\documentclass[10pt,aps,prb,twocolumn,amsmath,amssymb,floatfix,superscriptaddress]{revtex4-2}
\usepackage{graphicx}
\usepackage{bm}
\usepackage{bbm}
\usepackage{siunitx}
\usepackage[
colorlinks,
citecolor=blue,
linkcolor=blue,
urlcolor=blue,plainpages=false,pdfpagelabels
]{hyperref}
\usepackage[capitalize]{cleveref}
\usepackage{braket}
\usepackage{qcircuit}
\usepackage{lipsum}
\usepackage{comment}

\usepackage{xcolor}

\newcommand{\pd}[2]{\frac{\partial #1}{\partial #2}}
\newcommand{\pdd}[2]{\frac{\partial^2 #1}{\partial #2^2}}
\renewcommand{\v}[1]{\ensuremath{\mathbf{#1}}}

\DeclareMathOperator{\Tr}{Tr}
\newcommand{\half}{\frac{1}{2}}

\newcommand{\vb}[1]{\boldsymbol{\mathrm{#1}}}
\usepackage{derivative}
\newcommand{\mat}[1]{\begin{bmatrix}#1\end{bmatrix}}

\newcommand{\abm}[1]{{\color{red} ABM: #1}}
\newcommand{\fkm}[1]{{\color{blue} FKM: #1}}
\newcommand{\mjk}[1]{{\color{cyan} MJK: #1}}

\begin{document}

\title{
Functional matrix product state simulation of continuous variable quantum circuits
}

\newcommand{\affA}{Department of Computer Science, IT University of Copenhagen, DK-2300 Copenhagen S, Denmark.}
\newcommand{\affB}{Kvantify Aps, DK-2300 Copenhagen S, Denmark.}
\newcommand{\affC}{AWS Center for Quantum Computing, Pasadena, CA 91125, USA.}
\newcommand{\affD}{Department of Physics and Astronomy, Aarhus University, DK-8000 Aarhus C, Denmark}
\newcommand{\affE}{Department of Computer Science, University of Copenhagen, DK-2100 Copenhagen Ø, Denmark}
\newcommand{\affF}{NNF Quantum Computing Programme, Niels Bohr Institute, University of Copenhagen, DK-2100 Copenhagen Ø, Denmark}

\author{Andreas Bock Michelsen}
\affiliation{\affA}
\affiliation{\affB}
\affiliation{\affF}

\author{Frederik K. Marqversen}
\affiliation{\affB}
\affiliation{\affD}

\author{Michael Kastoryano}
\affiliation{\affE}
\affiliation{\affC}

\date{\today}

\begin{abstract}

We introduce a functional matrix product state (FMPS) based method for simulating the real-space representation of continuous-variable (CV)  quantum computation. This approach efficiently simulates non-Gaussian CV systems by leveraging their functional form. By addressing scaling bottlenecks, FMPS enables more efficient simulation of shallow, multi-mode CV quantum circuits with non-Gaussian input states. 
The method is validated by simulating random shallow and cascaded circuits with highly non-Gaussian input states, showing superior performance compared to existing techniques, also in the presence of loss.
\end{abstract}

\maketitle

\section{Introduction}

Continuous variable (CV) quantum systems have emerged as a promising platform for fault-tolerant quantum computing \cite{menicucci2006universal,larsen2021fault,bourassa2021blueprint,chamberland2022building,campagne2020quantum}, leveraging photonic or vibrational modes in quantum systems to encode qubits. These systems rely on quantum quadrature encodings, such as Gottesman-Kitaev-Preskill (GKP) states \cite{gottesman2001encoding} and cat states \cite{mirrahimi2014dynamically}, which serve as the building blocks for robust quantum information processing. These encodings, referred to as 'Bosonic qubits' \cite{ma2021quantum}, are reminiscent of classical phase-space encodings \cite{girvin2023introduction} and have been recognized for their potential in achieving high levels of fault tolerance in the quantum regime.

Bosonic qubits offer unique advantages over their discrete-variable counterparts. Chief among these is their intrinsic resilience to specific types of noise \cite{leghtas2013hardware,nathan2024self}. This is achieved through the physical structure of these states, which provide a natural layer of error correction at the hardware level. For instance, GKP states, which encode information into specific grid-like structures in phase space, are highly robust against small displacements, while cat states leverage coherent state superpositions that exhibit robustness against dephasing. These properties make CV systems not only theoretically attractive but also highly practical for scalable quantum computing architectures \cite{putterman2025hardware}.

However, the scaling of protocols involving non-Gaussian CV systems presents significant computational challenges. Non-Gaussianity is a cornerstone for universal quantum computation in CV systems, but it introduces a level of complexity that renders many existing simulation techniques inadequate. Numerical approaches that rely on Fock space representations with tensor networks \cite{oh2024classical,vinther2024variational} or Gaussian approximations, such as the Bosonic backend \cite{bourassa2021fast} implemented in platforms like Xanadu’s Strawberry Fields \cite{killoran2019strawberry}, often scale poorly when simulating quantum circuits with even a modest number of non-Gaussian states \cite{chabaud2021classical, chabaud2023resources}. This limitation becomes particularly pronounced in scenarios involving high levels of squeezing, displacement, and non-Gaussian operations, where the computational cost grows exponentially.

While tensor networks are a widely used  for simulating discrete-variable quantum circuits \cite{pan2022simulation,liu2022validating,vincentJet2022}, their performance for CV systems has been rather limited. The large state-space required to represent continuous variable systems, combined with the intricacies of handling non-Gaussian operations, leads to inefficient scaling and restricts their applicability for practical CV quantum circuits.

In this work, we propose a novel numerical method based on a real-space representation of CV states, termed 'functional matrix product state' (FMPS). Unlike traditional approaches, FMPS leverages the functional form of CV states directly in real space, providing a computationally efficient framework for simulating non-Gaussian CV systems. We demonstrate that this method significantly outperforms existing techniques, particularly for systems involving Bosonic qubits and operations of practical relevance. By addressing the scaling bottlenecks of current methods, our approach paves the way for more efficient simulation and design of CV quantum protocols.

Specifically, we develop a real-space tensor network simulation method for CV quantum circuits. The entanglement structure in CV systems is analyzed through its connection to the continuous Schmidt decomposition. The method involves discretising real space, with a focus on optimizing the computational bounding box, which must be adjusted following specific quantum operations. We also address the impact of noise and discuss the role of interpolation in the simulation. To demonstrate the functional tensor network method, we simulate (i) a random shallow circuit of beam splitters acting on multiple input states and (ii) a random cascaded circuit. For highly non-Gaussian input states, such as GKP states, our method outperforms existing simulation approaches which struggle to capture non-Gaussianity to high precision. This mirrors the behavior observed in discrete-variable systems, where tensor networks excel at simulating certain local shallow circuits. Finally, we extend the method to simulate noisy circuits, albeit with increased computational cost.

The methods developed in this paper have been applied in Ref. \cite{marqversenImpact2025} to study the effect of finite squeezing on the accuracy of measurement based quantum computing with GKP qubits, a work which illustrates how simulation of continuous variable systems can benefit from FMPS structure.





\section{The functional matrix product state decomposition}
\begin{figure}
    \centering
    \includegraphics[width=.8\columnwidth]{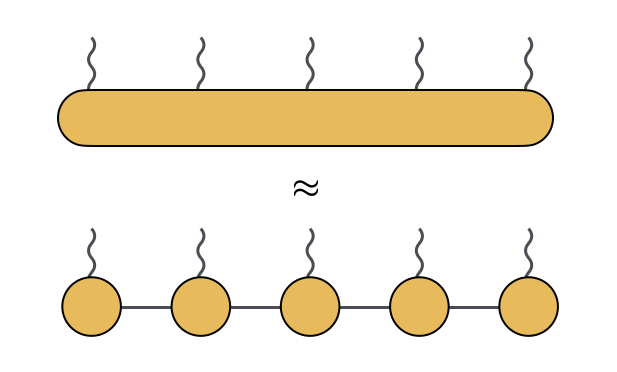}
    \caption{Functional matrix product state decomposition. Undulating lines represent continuous variables, while straight lines represent discrete variables. }
    \label{fig:diagram}
\end{figure}
We will be  considering multi-mode continuous variable quantum optics systems in the position basis:
\begin{align}
    \ket{\Psi} = \int d^m \v{q} f(\v{q}) \ket{\v{q}}.
\end{align}
The position space wave function $f(\v{q})$ is square-integrable with respect to the Lebesgue measure,
\begin{align}
    f(\v{q}) \in L^2(\mathbb{R}^m),
\end{align}
meaning we can apply the following results from Ref. \cite{bigoniSpectral2016}. 
The functional Schmidt decomposition can be applied to a general function of $m$ modes as
\begin{align}
    f(\v{q}) = \sum_{\alpha_1=1}^\infty \sqrt{\lambda_1(\alpha_1)} \gamma_1(q_1;\alpha_1) \varphi_1(\alpha_1;q_2,\dots,q_m).
\end{align}
We can then decompose $\varphi_1$ to find
\begin{multline}
    \sqrt{\lambda_1(\alpha_1)} \varphi_1(\alpha_1;q_2,\dots,q_m) \\
    = \sum_{\alpha_2=1}^\infty \sqrt{\lambda_2(\alpha_2)} \gamma_2(\alpha_1;q_2;\alpha_2) \varphi(\alpha_2;q_3,\dots,q_m).
\end{multline}
Repeating this process for each mode, yields the  FMPS decomposition 
\begin{align}
    f(\v{q}) = \sum_{\alpha_1,\dots,\alpha_{m-1}=1}^\infty \gamma_1(\alpha_0;q_1;\alpha_1) \cdots \gamma_m(\alpha_{m-1};q_m;\alpha_{m}),
\end{align}
where we have introduced the dummy indices $\alpha_0 = \alpha_m = 1$ for notational simplicity, and where we define the $m$'th function as
\begin{align}
    \gamma_m(\alpha_{m-1};q_m;\alpha_{m}) = \sqrt{\lambda_{m-1}(\alpha_{m-1})}\varphi_m(\alpha_{m-1};q_m).
\end{align}
Truncating each sum such that the sum over the index $\alpha_i$ only goes up to $r_i$, we find the finite FMPS approximation
\begin{align}
    f(\v{q}) \simeq \sum_{\alpha_1,\dots,\alpha_{m-1}=1}^\v{r} \gamma_1(\alpha_0;q_1;\alpha_1) \cdots \gamma_m(\alpha_{m-1};q_m;\alpha_{m}),
    \label{eq:FMPS_approx}
\end{align}
where $\v{r} = (r_1, \dots, r_{m-1})$ are the bond dimensions for each truncation. It is important to ensure that $\sum_i^r \lambda_i = 1$ to maintain normalisation of states. This decomposition is a natural extension of the discrete matrix product state (MPS) decomposition to continuous multimode states. The decomposition is shown diagrammatically in \cref{fig:diagram}, where $f(\v{q})$ is shown as a single object with several continuous variables (illustrated by undulating lines), and the decomposition in \cref{eq:FMPS_approx} is show as a series of tensors (circles) with one or two discrete indices (illustrated by straight lines) and one continuous variable.  

\subsection{Discretisation}
\label{sec:discretisation}
In the continuum, the eigenfunctions $\gamma_j (\alpha_j;q_j;\alpha_{j+1})$ are obtained by choosing an appropriate (product) measure, and evaluating the Schmidt decomposition in this measure space. In practice, we will evaluate the eigenfunctions by (i) truncating the space to a finite rectangular domain $I_{\textbf{q}}\subset \mathbb{R}^m$ and defining the uniform measure on this domain, and then (ii) a uniform discretisation of the space. In this way the FMPS quantum circuit simulation problem completely reduces to a discrete MPS simulation. The additional complications come from the need to (a) update the bounding box after every gate operation, and (b) ensuring that space discretisation is sufficiently refined to faithfully capture the computation. 

We will need to evaluate the wave functions off the grid points on which they are supported. For this we will apply cubic splines with Dirichlet boundary conditions (i.e. the functions have value and derivative equal to zero at the boundaries). The cubic spline scales approximately with the fourth power of the grid spacing and the fourth order derivative of the function being approximated.

An important concern is whether the smoothness of the discretised function is captured accurately, especially for highly oscillatory functions. In an FMPS context, a poorly captured function will be noisily described and lose important features such as smoothness and symmetry, which translates into a function that is difficult to describe in the FMPS framework. In the extreme case of a function that is pure noise, the bond dimension required to accurately capture the function would be equal to the number of discretisation points $N$, while for a faithful representation, the bond dimension would be independent of $N$.

\begin{figure}
    \centering
    \includegraphics[width=0.85\columnwidth]{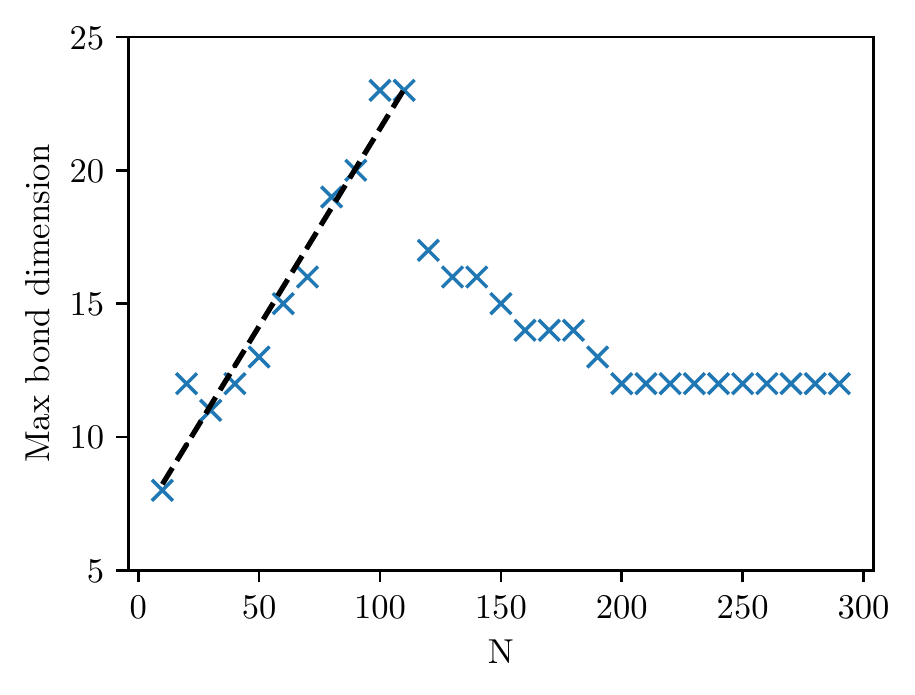}
    \caption{Simulating a 10-step cascaded circuit of cat states with a gate fidelity of $0.999$ and interpolation with cubic splines, we see that the maximum bond dimension  increases linearly with the number of grid points $N$ for a while, before it falls back down and converges. Once the bond dimension no longer changes with $N$, we have a faithful representation of the state. The dashed line is included as a guide to the eye.}
    \label{fig:critical-N}
\end{figure}

In practice, we need to ensure that $N$ is large enough such that the representation of the function is unaffected by small changes to $N$. This can be tested by checking whether the bond dimension changes with $N$. An example is shown in \cref{fig:critical-N}, with details of the circuit involved described in \cref{sec:cascade}. For a fixed precision, the required bond dimension initially grows linearly, before it stabilizes down to a constant value reflecting the singular value truncation of the continuous system.  Interestingly, for intermediate values of $N$, the bond dimension grows significantly larger than the final (large $N$) limit, reflecting a signal to noise tradeoff. 

\subsection{Implications of discretisation on momentum space}
As we work with a discretised position space, it is important to consider the implications of  discretisation in momentum space as well. Otherwise, we might get an unfaithful representation of the wave function. In this section, we investigate these implications by use of the theory from Ref. \cite{oppenheim_discrete-time_2010}. Note, that the momentum space wave function is obtained through a Fourier transform, or equivalently by rotating by $-\pi/2$ through application of $\hat{P}(-\pi/2)$ (see Eqn, (\ref{eqn:phaserot}).

Let $I_q \subset \mathbb{R}$ be a finite interval containing the support in position space of the wave function $\psi$ such that 
\begin{equation}
    \psi(q) \approx 0 
        \quad
    \forall q \not\in I_q
\end{equation}
If the number of sample points in position is $N$, then the sample rate  is $r_q = N / |I_q|$, where $|I_q|$ is the length of the interval. The Nyquist theorem guarantees no aliasing in momentum space when
\begin{equation}
    |I_p| \leq 2\pi r_q.
\end{equation}
In other words, given that the width of $\psi$ in momentum space is $|I_p|$, then the minimal sample rate in position $r_q$ which will guarantee a faithful representation is given by the above inequality. Note, that the sample rate cannot simply be increased by interpolation, since the choice of interpolation scheme significantly impacts the momentum space representation. 

Choosing a sample rate as determined by the Nyquist theorem has the additional advantage that the full position space wave function can be reconstructed \emph{exactly} by sinc-interpolation. However, usually wave functions will not have finite support. In fact it is impossible for a function to have finite support in both position and momentum space simultaneously. Sinc-interpolation thus is only a good approximation, and in practice we observe that the sampling rates are usually large enough that the difference between sinc-interpolation and something highly optimised like cubic spline interpolation is negligible.

\subsection{Accuracy}
In total, simulating CV states in the discretised FMPS formalism involves three approximations, namely:
\begin{itemize}
	\item Trunctation to a bounding box.
    \item Discretisation of position or momentum space.
    \item Truncation to a finite MPS representation.
\end{itemize}
Managing errors due to discretisation is discussed in the previous section, and in practice a manageable value of $N\sim 200$ along with cubic spline interpolation leads to small errors in the experiments that we have run. Higher accuracy with lower $N$ might be reached with adaptive, non-regular grids, but this is beyond the scope of our work.

The error due to truncation of the MPS representation can be better controlled, since the loss of fidelity in a truncated state is given by the total singular value weight lost in the truncation,
\begin{align}
    \mathcal{F}_\text{trun} = 1 - \sum_{j=r+1}^N \sigma_j,
\end{align}
where $r$ is the bond dimension and $\sigma_j$ is the $j$'th normalised singular value. It is thus simple to implement a strategy where the chosen bond dimension is controlled by a target fidelity. 

Given the fidelity of each applied gate in a circuit, the total fidelity of the end state $\mathcal{F}$ can be estimated as \cite{ayralDensityMatrix2023,olivaEntanglementaware2023}
\begin{align}
    \mathcal{F} \approx \prod_{j=1}^G \mathcal{F}_{j}^{1/G},
\end{align}
where $\mathcal{F}_{j}$ is the fidelity of the $j$'th gate, and $G$ is the number of gates applied. However, this can only ever be an accurate estimate as long as the discretisation error is negligible.

\subsection{Scaling}
The benefit of the FMPS formalism is that it inherits the scaling of MPS methods. Consider a circuit of $m$ modes represented on a discretised spatial grid of $N$ points (known as the physical dimension of the tensor) with $G$ two-mode nearest neighbour gates applied and a maximum bond dimension $r_\text{max}$. We can approximate any state in this system with $\mathcal{O}(m N r_\text{max}^2)$ elements. The primary computational cost comes from performing the $G$ SVD's. Performing a SVD on an $N_1 \times N_2$ matrix scales as $\mathcal{O}(N_1 N_2^2)$ for $N_1\geq N_2$. This means that after a two-mode gate resulting in a tensor core with physical dimensions $(N_1,N_2)$ and outer bond dimensions $r_1,r_2$, we can return to MPS form by performing SVD which scales as $\mathcal{O}(r_1 N_1 [r_2 N_2]^2)$. Finally, evaluating expectation values involves fast MPS contractions and the evaluation of $r_1 \times \cdots \times r_m$ one-dimensional integrals. 

As long as the bond dimension is bounded, representing the state requires less than exponential resources in all parameters. The trade-off is the approximate representation and the worse scaling in number of gates, which each cost an SVD, as well as a poor ability to represent highly entangled states, which require an unfeasible bond dimension. The sub-exponential scaling in the number of modes constitutes a central result of our work. 

\section{Useful operators}

We consider the transformation of a discretised CV state under the action of some of the most important CV quantum operations. First we analyse the single mode displacement, squeezing, and phase rotation operators. We also analyse the two-mode rotation operator, also known as the beam splitter. These form a basis for all multi-mode Gaussian operators that are important for CV quantum mechanics. Finally we also include the non-Gaussian cubic phase gate. All-together, these operators are sufficient for universal CV quantum computing \cite{weedbrookGaussian2012}. In each case, we indicate how the bounding box (see \cref{fig:bounding_box} and \cref{tab:domain-summary}) needs to be modified to preserve the computation in the bounded domain. Certain operations can incur increased requirements for discretisation $N$. However, we found that it is more convenient to leave $N$ as a free global parameter in simulations. 

In the following sections, the discretised CV state wave function will be denoted $\psi(\vb{q})$ with $\vb{q} \in I \subset \mathbb{R}^n$ an interval containing its support (approximately). The derivations given below rely mostly on the knowledge of the operators' symplectic forms. For further details on this subject, the authors refer to Ref. \cite{weedbrookGaussian2012}.

\begin{table}[h]
    \centering
    \begin{tabular}{l|c|c}
         Operation & Operator & Domain effect\\
         \hline
         Displacement & $e^{i (d_2 \hat{q} - d_1 \hat{p})}$ & $I \to I + d_1$\\
         Squeezing & $e^{-i\frac{s}{2} (\hat{q}\hat{p} + \hat{p}\hat{q})}$ & $I \to e^s I$\\
         Phase rotation & $e^{i\frac{\phi}{2} (\hat{q}^2 + \hat{p}^2 - 1)}$ & $I \to (\langle q \rangle - n \sigma, \langle q \rangle + n \sigma)$\\
         Beam splitting & $e^{\theta(\hat{q}_0 \hat{p}_1 - \hat{p}_0 \hat{q}_1)}$ & $I_j \to (\min_k\{p'^k_j\}, \max_k\{p'^k_j\})$ \\
         Cubic phase gate & $e^{i \frac{\gamma}{6} \hat{q}^3}$ & $I \to I$
    \end{tabular}
    \caption{A summary of the effects of each operation on the domain. }
    \label{tab:domain-summary}
\end{table}

\subsection{Displacement}
The quadrature displacement operator is given by
\begin{equation}
    \hat{D}(\vb{d}) = e^{i (d_2 \hat{q} - d_1 \hat{p})}
        ,\quad
    \hat{D}(\vb{d})^\dagger \mat{\hat{q} \\ \hat{p}} \hat{D}(\vb{d}) = \mat{\hat{q} + d_1 \\ \hat{p} + d_2}
\end{equation}
where $\vb{d} = (d_1, d_2)^T \in \mathbb{R}^2$. The action of the displacement operator on the $q$-quadrature eigenstates is the phase space displacement
\begin{gather}
    \hat{D}(\vb{d}) \ket{q} \propto e^{i d_2 q} \ket{q + d_1}
\end{gather}
where proportionality is up to an irrelevant global phase (constant over all $q \in \mathbb{R}$). Acting on a quantum state this becomes
\begin{equation}
    \hat{D}(\vb{d}) \ket{\psi} = \int_I \odif{q} \psi(q) \hat{D}(\vb{d})\ket{q}
    \propto \int_{(I + d_1)} \odif{q} e^{i d_2 q} \psi(q - d_1) \ket{q}
\end{equation}
again up to global phase. From this we read off the wave function transformation
\begin{align}
    \psi(q) \to e^{i d_2 q} \psi(q-d_1)
        ,\quad 
    I \to I + d_1
\end{align}

\subsection{Squeezing}
The $p$-quadrature squeezing operator is given by
\begin{equation}
    \hat{S}(r) = e^{\frac{s}{2} (\hat{a}^{\dagger 2} - \hat{a}^2)} 
    = e^{-i\frac{s}{2} (\hat{q}\hat{p} + \hat{p}\hat{q})}
\end{equation}
and has symplectic representation
\begin{equation}
    \hat{S}(r)^\dagger \mat{\hat{q} \\ \hat{p}} \hat{S}(s) = \mat{e^s & 0 \\ 0 & e^{-s}} \mat{\hat{q} \\ \hat{p}}
\end{equation}
such that $s>0$ corresponds to $p$-squeezing and $q$-broadening. Applying the squeezing operator has the overall effect of stretching the coordinate axis of the corresponding mode.
\begin{equation}
    \hat{S}(s) \ket{q} \propto e^{\frac{s}{2}} \ket{e^s q}
\end{equation}
where proportionality again is up to global phase. The factor $e^{\frac{s}{2}}$ is the required normalisation. General states thus transform as
\begin{equation}
    \hat{S}(s) \ket{\psi} \propto \int_{e^s I} \odif{q} e^{-\frac{s}{2}} \psi(e^{-s} q) \ket{q},
\end{equation}
from which the following wave function transformation is read off
\begin{align}
    \psi(q) \to e^{-\frac{s}{2}} \psi(e^{-s} q)
        ,\quad 
    I \to e^s I
\end{align}

\subsection{Phase rotation}
A phase rotation in position space is implemented by the operator
\begin{align}
    \hat{P}(\phi) = e^{i \phi a^\dagger a} = e^{i\frac{\phi}{2} (\hat{q}^2 + \hat{p}^2 - 1)},\label{eqn:phaserot}
\end{align}
The position space representation corresponds to the Feynman propagator of the harmonic oscillator \cite[sec. 2.6]{sakurai_modern_2017}:
\begin{multline}
    \bra{q} \hat{P}(\phi) \ket{q'} = (2 \pi |\sin\phi|)^{-\half}
        \\
    \times \exp\left( \frac{-i}{\sin \phi} \left[\half (q^2+(q^\prime)^2)\cos\phi - q q^\prime \right] \right).
        \label{eq:mehler}
\end{multline}
up to an irrelevant global phase. The wave function transformation is given by
\begin{equation}
    \psi(q) \to \psi'(q) = \int dq' \psi(q') \bra{q} \hat{P}(\phi) \ket{q'}.
\end{equation}

Since the operator performs a rotation in phase space, the bounding box must also be transformed. The first moment (mean position) of the transformed wave function $\psi'$ can be calculated from the initial non-transformed wave function $\psi$ as
\begin{equation}
    \langle q \rangle_{\psi'} = \int dq \psi^*(q) (q \cos\phi + i \sin\phi \pd{}{q}) \psi(q),
        \label{eq:first moment}
\end{equation}
and for the second moment
\begin{multline}
    \langle q^2 \rangle_{\psi'} = \int dq \psi^*(q) \Bigg[ \cos^2\phi q^2 + i \cos\phi \sin\phi \left(1 + 2q \pd{}{q}\right) 
        \\
    \qquad - \sin^2\phi \pdd{}{q}\Bigg] \psi(q).
        \label{eq:second moment}
\end{multline}
Derivations of these identities are included in appendix \ref{app:moments}. From the first and second moments the variance can be computed
\begin{equation}
    \sigma_{\psi'}^2 = \langle q^2 \rangle_{\psi'} - \langle q \rangle_{\psi'}^2.
\end{equation}
This serves as an estimate on the width of the bounding box. In practice, one would have to choose some number $n \in \mathbb{N}$ of ``sigmas'' to use for the actual width of the bounding box. Together these give
\begin{align}
    I \to (\langle q \rangle - n \sigma, \langle q \rangle + n \sigma).
\end{align}
Whether or not the right number of sigmas has been chosen can be determined by asserting normalisation of the state.

An alternative strategy is to keep track of both $p$ and $q$ through the hyperbox method described at the end of the next section.

\subsection{Beam splitter}
The beam splitter acting on modes $0$ and $1$ is given as
\begin{gather}
    \hat{R}(\theta) = e^{i\theta(a_0^\dagger a_1 - a_0 a_1^\dagger)} = e^{\theta(\hat{q}_0 \hat{p}_1 - \hat{p}_0 \hat{q}_1)}
\end{gather}
where subscripts refer to the mode an operator acts on. The beam splitter has symplectic action
\begin{equation}
    \hat{R}(\theta)^\dagger \mat{\hat{q_0} \\ \hat{q_1}} \hat{R}(\theta) = \mat{\cos\theta & -\sin\theta \\ \sin\theta & \cos\theta} \mat{\hat{q_0} \\ \hat{q_1}}
    \label{eq:symplectic_rotation}
\end{equation}
and the same for the $p$-quadratures. That is, the action of the beam splitter is a simple rotation of the quadratures. Writing $\vb{q} = [q_0, q_1]^T$ and letting $O$ denote the orthogonal matrix in \cref{eq:symplectic_rotation} this gives
\begin{equation}
    \hat{R}(\theta) \ket{\vb{q}} = \ket{O \vb{q}}.
\end{equation}
Using orthogonality: $O^T O = I$ and $\det{O} = 1$, this generalises to any two-mode quantum state as:
\begin{equation}
    \hat{R}(\theta) \ket{\psi} = \int_I \odif[ord=2]{\vb{q}} \psi(\vb{q}) \ket{O \vb{q}}
    = \int_{OI} \odif[ord=2]{\vb{q}} \psi(O^T \vb{q}) \ket{\vb{q}},
\end{equation}
which gives the wave function transformation
\begin{equation}
    \psi(\vb{q}) \to \psi(O^T \vb{q})
\end{equation}
Even if the initial domain is axis-aligned $I = I_0 \times I_1$, the transformed domain $I' = OI$ will in general not be so. In order to apply the Schmidt decomposition to the two modes, an axis-aligned interval is required.

\begin{figure}
    \centering
    \includegraphics[width=\linewidth]{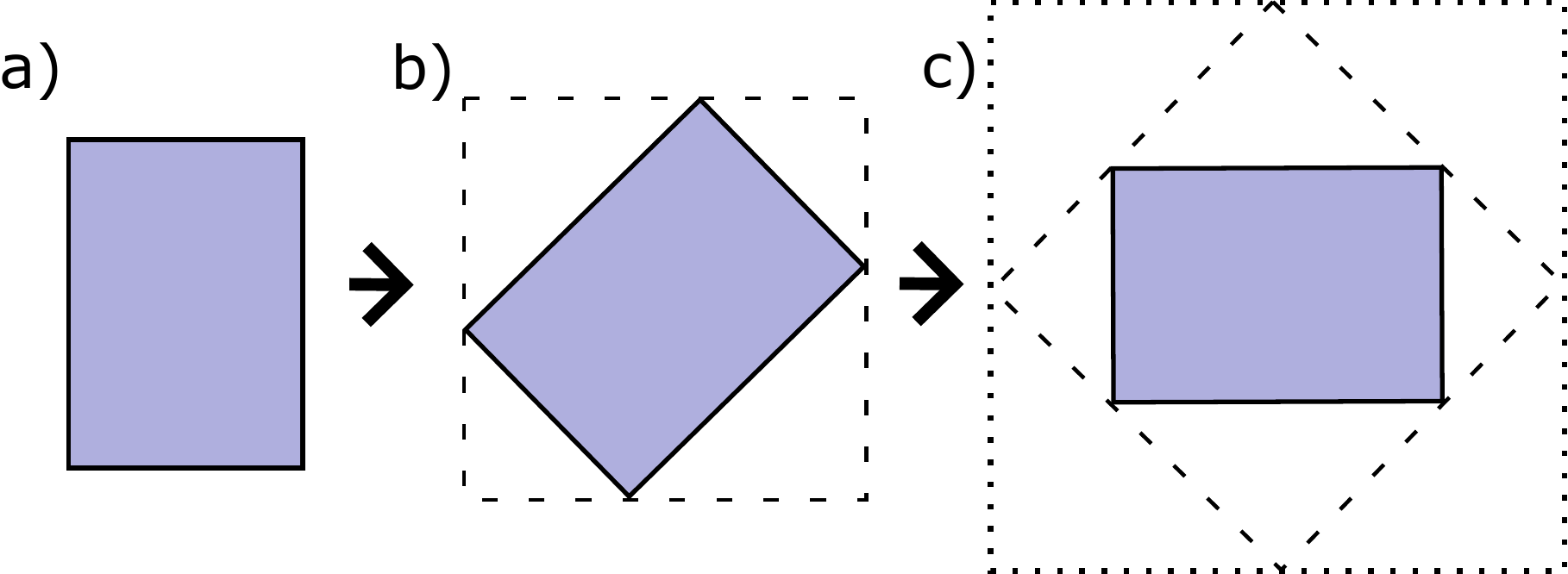}
    \caption{When rotating the domain of our state, it is important that we keep track of the original domain such that the bounding box does not grow unnecessarily large upon repeated rotations. Here this is illustrated with a two-dimensional domain undergoing two $\pi/4$ rotations. At step \textit{b}, the dashed bounding box defines the new domain, but further rotating that domain to step \textit{c} results in an unnecessarily large domain (dotted box), mostly without information of the state. Instead, keeping track of the original domain lets us shrink the domain in step \textit{c} to the solid line.}
    \label{fig:bounding_box}
\end{figure}

One option is to choose the axis-aligned bounding box that contains the entirety of $I'$. Let $\vb{p}^k$ denote the four corners of $I$. The corners of $I'$ are then simply $\vb{p}'^k = O \vb{p}^k$, and the sought after new domain becomes
\begin{align}
    I_0 \to (\min_k\{p'^k_0\}, \max_k\{p'^k_0\}),
        \\
    I_1 \to (\min_k\{p'^k_1\}, \max_k\{p'^k_1\}).
\end{align}
Keeping the number of grid points in the discretisation constant will in general imply a reduced resolution of the transformed state due to the volume of the bounding box being greater than that of $I'$. Alternatively the resolution can be preserved by increasing the number of points such that the density along each axis is kept constant. In the case of consecutive rotations, the procedure should be repeated for the original domain $I$, otherwise the bounding box would grow unnecessarily large, see \cref{fig:bounding_box}. In practice, this can be implemented by assigning an $m$-dimensional hyperbox to the initial state, and then updating it along with the states under the application of operations. The description of such a hyperbox requires $2^m$ numbers, so for a very large number of modes, alternative strategies might be preferable.

\subsection{Cubic phase gate}
The cubic phase gate is given as
\begin{equation}
    \hat{C}(\gamma) = e^{i \frac{\gamma}{6} \hat{q}^3}
        ,\quad
    \hat{C}(\gamma)^\dagger \begin{bmatrix}\hat{q} \\ \hat{p}\end{bmatrix} \hat{C}(\gamma) = \begin{bmatrix}\hat{q} \\ \hat{p} + \gamma \hat{q}^2\end{bmatrix}
\end{equation}
Being diagonal in the $q$-quadrature, this operator has the following almost trivial action on the eigenstates
\begin{equation}
    \hat{C}(\gamma) \ket{q} = e^{i \frac{\gamma}{6} q^3} \ket{q}
\end{equation}
from which it follows that the wave function transformation becomes
\begin{align}
    \psi(q) \to e^{i \frac{\gamma}{6} q^3} \psi(q)
        ,\quad 
    I \to I
\end{align}

\section{Useful measurements}
\label{sec:measurements}

This section builds on theory of bosonic measurements from Ref. \cite{weedbrookGaussian2012}.

\subsection{Photon number measurement}

Consider a general position space state with $m$ modes, $\psi(q_1,\dots,q_m)$. The probability distribution of a photon number measurement in mode $i$ yielding the result $n$ is
\begin{align}
    p_i(n) = \Tr(\rho \ket{n,i}\bra{n,i})
\end{align}
where $\rho$ is the density matrix which for a pure state is simply $\ket{\psi}\bra{\psi}$, and $\ket{n,i}$ is the Fock state describing $n$ photons in mode $i$. In position space, we have
\begin{multline}
    p_i(n) = \int d^m\v{q}\ d^m\v{q}^\prime\ \psi^*(\v{q}) \psi(\v{q}^\prime) \\ \times \braket{q_1|q_1^\prime} \cdots \braket{n,i|q_i}\braket{q_i^\prime|n,i}\cdots \braket{q_m|q_m^\prime}.
\end{multline}
By evaluating the inner products and recalling the functional tensor train form of $\psi(\v{q})$ given in \cref{eq:FMPS_approx}, we find
\begin{multline}
    p_i(n) \propto \sum_{\alpha,\alpha'} \int dq_1 \gamma_1(\alpha_0;q_1;\alpha_1) \gamma^*_1(\alpha_0';q_1;\alpha_1') \cdots\\ 
    \times \int dq_i\ h_n(q_i) \gamma_i(\alpha_{i-1};q_i;\alpha_{i}) \int dq_i^\prime\ h_n(q_i^\prime) \gamma^*_i(\alpha_{i-1}';q_i^\prime;\alpha_{i}') \\
    \cdots \times \int dq_m \gamma_m(\alpha_{m-1};q_m;\alpha_m) \gamma_m^*(\alpha_{m-1}';q_m;\alpha_m'),
    \label{eq:number_prob_distribution}
\end{multline}
where $h_n(q)$ is the Hermite function describing the position space Fock state
\begin{align}
    h_n(q) = \braket{q | n, i} = \frac{1}{\sqrt{2^n n! \sqrt{\pi}}}e^{-q^2/2} H_n(q),
\end{align}
with $H_n$ the $n$'th Hermite polynomial. Here we note a major benefit gained by working with FMPS's, namely that the resulting state is easily integrated over. In particular, \cref{eq:number_prob_distribution} involves $r_1 \times \dots \times r_m$ terms with $m$ one-dimensional integrals each, where $r_i$ is the bound dimension of the index $\alpha_i$. Assuming a numerical grid of $N$ points and $r_i \ll N$ for all $i$, this scales much better with the number of modes $m$ than the full integral over $N^m$ points, which quickly becomes intractable for any number of grid points large enough to approximate the continuum.

\subsection{Homodyne detection}
\label{subsec:homodyne}
A homodyne measurement projects onto the quadrature basis, with the probability of measuring $\tilde q$ in mode $i$ given by
\begin{align}
    p_i(\tilde q) = \Tr(\rho \ket{\tilde q_i}\bra{\tilde q_i})
\end{align}
In position space we have
\begin{multline}
    p_i(\tilde q) = \int d^m\v{q}\ d^m\v{q}^\prime\ \psi^*(\v{q}) \psi(\v{q}^\prime) \\ \times \braket{q_1|q_1^\prime} \cdots \braket{\tilde q_i|q_i}\braket{q_i^\prime|\tilde q_i}\cdots \braket{q_m|q_m^\prime}.
\end{multline}
The position space eigenfunction is the delta function $\braket{q | q'} = \delta(q - q')$, and so in FMPS form this becomes
\begin{multline}
    p_i(\tilde q) \propto \sum_{\alpha,\alpha'} \int dq_1 \gamma_1(\alpha_0; q_1; \alpha_1) \gamma^*_1(\alpha_0'; q_1; \alpha_1') \cdots
        \\ 
    \times \Big[ \gamma_i(\alpha_{i-1}; \tilde q_i; \alpha_{i}) \gamma^*_i(\alpha_{i-1}'; \tilde q_i^\prime; \alpha_{i}') \Big]
        \\
    \cdots \times \int dq_m \gamma_m(\alpha_{m-1};q_m;\alpha_m) \gamma_m^*(\alpha_{m-1}';q_m;\alpha_m'),
\end{multline}
where only the $i$'th integral has been resolved. To apply a homodyne measurement of the $p$ quadrature (or indeed any phase rotated quadrature), one can rotate that quadrature onto the $q$ quadrature, after which the homodyne measurement is simply performed as described above.

\subsection{Heterodyne detection}
A heterodyne measurement projects onto the coherent state basis, with the probability of measuring the coherent state given by $\vb{s} \in \mathbb{R}^2$ in mode $i$ given by
\begin{align}
    p_i(\vb{s}) = \Tr(\pi^{-1/2} \rho \ket{\vb{s},i} \bra{\vb{s},i}).
\end{align}
In position space we have
\begin{multline}
    p_i(\vb{s}) = \int d^m\v{q}\ d^m\v{q}^\prime\ \psi^*(\v{q}) \psi(\v{q}^\prime) \\ \times \braket{q_1|q_1^\prime} \cdots \braket{\vb{s},i|q_i}\braket{q_i^\prime|\vb{s},i}\cdots \braket{q_m|q_m^\prime}.
\end{multline}
where the position space representation of the coherent state $\vb{s} = [s_1, s_2]^T$ is 
\begin{equation}
    \braket{q | \vb{s}} 
        =
    \bra{q} \hat{D}(\vb{s}) \ket{0}
        =
    \pi^{-\frac{1}{4}} e^{\frac{i}{2} s_1 s_2} e^{-\half (q - s_1)^2 + i s_2 q}
\end{equation}
and so in FMPS form this becomes
\begin{multline}
    p_i(\beta) \propto \sum_{\alpha,\alpha'} \int dq_1 \gamma_1(\alpha_0;q_1;\alpha_1) \gamma_1^*(\alpha_0';q_1;\alpha_1') \cdots\\
    \times \int dq_i\ \braket{q_i | \vb{s}, i}^* \gamma_i(\alpha_{i-1};q_i;\alpha_{i}) \\
    \times \int dq_i^\prime\ \braket{q_i^\prime | \vb{s}, i} \gamma^*_i(\alpha_{i-1}';q_i^\prime;\alpha_{i}')\\
    \cdots \times \int dq_m \gamma_m(\alpha_{m-1};q_m;\alpha_m) \gamma_m^*(\alpha_{m-1}';q_m;\alpha_m').
\end{multline}


\section{Noise}
\label{sec:noise}

\begin{figure}
    \centering
    \includegraphics[width=.8\columnwidth]{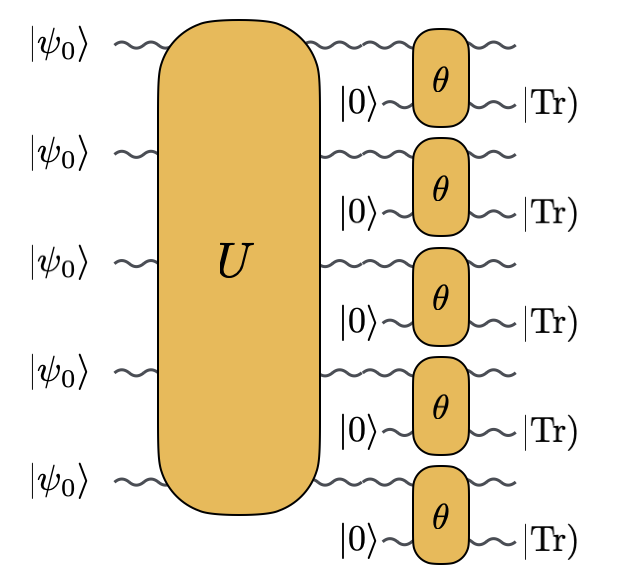}
    \caption{Noisy simulations. The loss noise is commuted through to the end of the computation. The trace can naturally be applied to evaluate expectation values in the tensor network framework, by contracting the indices with their conjugate copy.  }
    \label{fig:noise-diagram}
\end{figure}

To perform simulations relevant to experimental situations, it is critical to be able to predict the effects of noise. This can be done with the FMPS method while main reasonable scaling using the following strategy. We  focus on photon loss, the primary source of noise in photonic quantum computing. Other noise sources can also be treated in similar ways (e.g., phase noise, partial photon distinguishability, fabrication imperfections, and detector dark counts) \cite{shchesnovichNoise2019,qiRegimes2020}. Our implementation leverages the fact that uniform photon loss commutes with linear optical elements, including all two-mode gates used in CV quantum computing \cite{liuSimulating2023}. As a result, 
the loss noise can be commuted through the circuit, and only simulated at the end of the computation. That way, we can separate the noisy simulation into a noiseless quantum circuit simulation followed by a single round of noisy (loss) channel simulation, see \cref{fig:noise-diagram}, as is customary in Gaussian circuit simulations \cite{oh2024classical}. 

An accurate representation of the lossy state would normally require integrating out the vacuum modes and thus moving from a wave function description to a density matrix description. In the FMPS formalism, however, this transition significantly increases the scaling. Since the density matrix is typically not the primary object of interest—rather, one aims to calculate an observable derived from it—it is often more natural to work in the purification picture, and trace out the environment modes at the stage of computing expectation values. This approach lends itself to two key strengths of the formalism: easily adding or removing modes, and rapidly performing tensor contractions for expectation values. 


As an example, we consider the $q$-quadrature of a mode $i$ of a circuit as discussed in \cref{subsec:homodyne}.  Given the pure state
\begin{align}
    \ket{\psi} = \int d^m \v{q} \sum_{\alpha}^\v{r} \gamma_1(\alpha_0;q_1;\alpha_1) \cdots \gamma_m(\alpha_{m-1};q_m;\alpha_{m}) \ket{\v{q}},
\end{align}
we can beam split with the vacuum
\begin{multline}
    \hat R(\theta) \int dq_{j,a} dq_{j,b} \gamma_j(\alpha_{j-1};q_{j,a};\alpha_{j}) h_0(q_{j,b}) \ket{q_{j,a},q_{j,b}}
    \\ = \int dq_{j,a} dq_{j,b} f(\alpha_{j-1};q_{j,a},q_{j,b};\alpha_1) \ket{q_{j,a},q_{j,b}}
\end{multline}
where $a$ indicates the circuit subsystem, $b$ indicates the vacuum subsystem, $h_0(q)$ is the vacuum position space wave function and $f$ is some function. We can then perform the Schmidt decomposition
\begin{multline}
    f(\alpha_{j-1};q_{j,a},q_{j,b};\alpha_1) \\=
    \sum_\beta  \chi_{j,a}(\alpha_{j-1};q_{j,a};\beta_{j}) \chi_{j,b}(\beta_{j};q_{j,b};\alpha_{j})
\end{multline}
to get back to FMPS form. Hence, the total cost of adding noise reduces to a bounded number of decompositions, at most $m r_\text{max}^2$ for $m$ modes with maximum bond dimension $r_\text{max}$, and contracting a tensor network of twice the original length when extracting expectation values. Beam splitting each mode with the vacuum and performing the decomposition can be parallelised, as it is independent of the rest of the circuit.

\section{Examples}

Here, we demonstrate that the FMPS method is fast and robust for shallow Gaussian circuits with highly non-Gaussian input states. In particular we compare to the \texttt{bosonic} backend of the photonics library Strawberry Fields (SF) \cite{bourassaFast2021} with default settings unless otherwise noted. Whenever both an FMPS and an SF solution are available, we compare them by calculating the quadrature distance
\begin{align}
    \varepsilon = \sum_{k=0}^m \sum_{j=0}^N \left| |\psi_{k,\text{SF}}(q_j)|^2 - |\psi_{k,\text{FMPS}}(q_j)|^2 \right|,
\end{align}
where the inner sum is over each point on the $q$-quadrature grid, and the outer sum is over each mode. This measure is chosen since the $q$-quadrature calculation is a built-in method in SF and is easily done with FMPS since it involves simple tensor contraction. 

While the FMPS method works well for general non-Gaussian input states, we will be considering three representative classes for the examples: squeezed states, cat states and GKP states.
The squeezed vacuum state (or simply squeezed state), which is a Gaussian minimum-uncertainty state with uneven $q$- and $p$-quadratures, has the position wave function
\begin{align}
    \braket{q|s} = \mathcal{N} \exp\left[-\half \left(\frac{q}{e^{s}}\right)^2 \right],
\end{align}
where $s$ is the squeezing parameter and $\mathcal{N}$ is normalisation. The non-Gaussian cat state is the sum of two coherent states,
\begin{align}
    \ket{\psi}_\text{cat} &\propto \ket{\alpha} +e^{i\theta}\ket{-\alpha},\label{eqn:cat}
\end{align}
where $\alpha$ is the displacement amplitude of the coherent state. For a general displacement $\alpha \in \mathbb{C}$ the position wave function of the cat state is
\begin{align}
    \braket{q|\alpha} = \mathcal{N} e^{-\half [q-\sqrt{2}\text{Re}(\alpha)]^2+ i \text{Im}(\alpha) q},
\end{align}
with $\mathcal{N}$ a normalisation constant.
Finally, the finite energy GKP state is the superposition of the logical one and zero states
\begin{align}
    \ket{\psi}_\text{gkp} \propto \cos \frac{\theta}{2} \ket{0}_\text{gkp} + e^{-i\phi} \sin \frac{\theta}{2} \ket{1}_\text{gkp},\label{eqn:gkp}
\end{align}
which is a non-Gaussian state as well. The position wave functions of the logical states are defined \cite{tzitrinProgress2020} as a series of Gaussians at $2 \sqrt{\pi}$ intervals under a Gaussian envelope,
\begin{align}
    \braket{q|0}_\text{GKP} &= \mathcal{N} e^{-\frac{(\kappa q)^2}{2}} \sum_{n=-\infty}^{\infty} e^{-\frac{1}{2 \Delta^2} [2n \sqrt{\pi}]^2},\\
    \braket{q|1}_\text{GKP} &= \mathcal{N} e^{-\frac{(\kappa q)^2}{2}} \sum_{n=-\infty}^{\infty} e^{-\frac{1}{2 \Delta^2} [(2n+1) \sqrt{\pi}]^2},
\end{align}
where $\kappa$ gives the width of the envelope, and $\Delta$ gives the width of the Gaussians under the envelope.

\subsection{Numerics}
In the following, all physical parameters have been chosen at random within an interval to avoid engineering the circuits to be particularly advantageous or disadvantageous for the methods involved. Complex or real numbers will have their amplitude chosen randomly inside a given interval, while their phases, and angles in general, will be chosen randomly over the interval $[0,2\pi]$. These choices are made once for each circuit, and remain the same as circuits scale up in number of modes. 

To assess the efficiency of the respective methods, wall times for the computations are reported. These include initial state preparation and quantum circuit evolution. They do not include the time it would take to re-express the result of one method in the preferred format of the other method. The precision of the timing  involved is $\sim 1$ ms, and any measurements of $0$ ms have been corrected to $1$ ms. This correction does not alter the interpretation of  the data.

For numerical comparison, we chose settings for the FMPS implementations which are as simple as possible, while staying robust to discretisation/truncation errors and yielding negligible errors in comparison with SF implementations at default settings. The specifications are given in Table II. The timing and accuracy of the SF simulations are also under default settings. We found that non-standard settings which reduced the precision of the SF method, such as weak Gaussian truncations, did not significantly impact the scaling of the method, as compared to the FMPS method. Whenever a comparison was possibly between the SF and FMPS results, the quadrature error was $\varepsilon \sim 10^{-6}$.

In some cases, the SF method will be stopped after running for $> 10^{3}$ seconds or if 32 GB of RAM proves to be insufficient memory. Simulations are run for five independent random instances, with the solid, dashed or dotted line denoting the average wall time, and the shaded regions the min and max wall time for each setting. 

Measurements have not been included in the comparisons since the Strawberry Fields library does not natively support photon number measurements in the chosen backend, and because circuit evolution is a bigger bottleneck of the FMPS method than measurements, as discussed in \cref{sec:measurements}.

\begin{table}
    \centering
    \label{tab:FMPS-wide-settings}
    \caption{FMPS settings in both brick wall and cascaded circuit simulations.}
    \begin{tabular}{c|ccccc}
                & Amplitude & $N$       & $q$-range & $\mathcal{F}_j$ & $r$ bound \\
        \hline
        Squeezed &  $|s| \leq 0.5$ & 200 & $[-5,5]$ & 0.99 & 50 \\
        Cat     & $|\alpha| \leq 2$  & 200 & $[-8,8]$ & 0.99 & 50 \\
        GKP     & $\kappa = \Delta = 0.453$ & 200 & $[-10,10]$ & 0.99 & 40\\
    \end{tabular}
\end{table}

\subsection{Two squeezed modes}

\begin{figure}
    \centering
    \includegraphics[width=\linewidth]{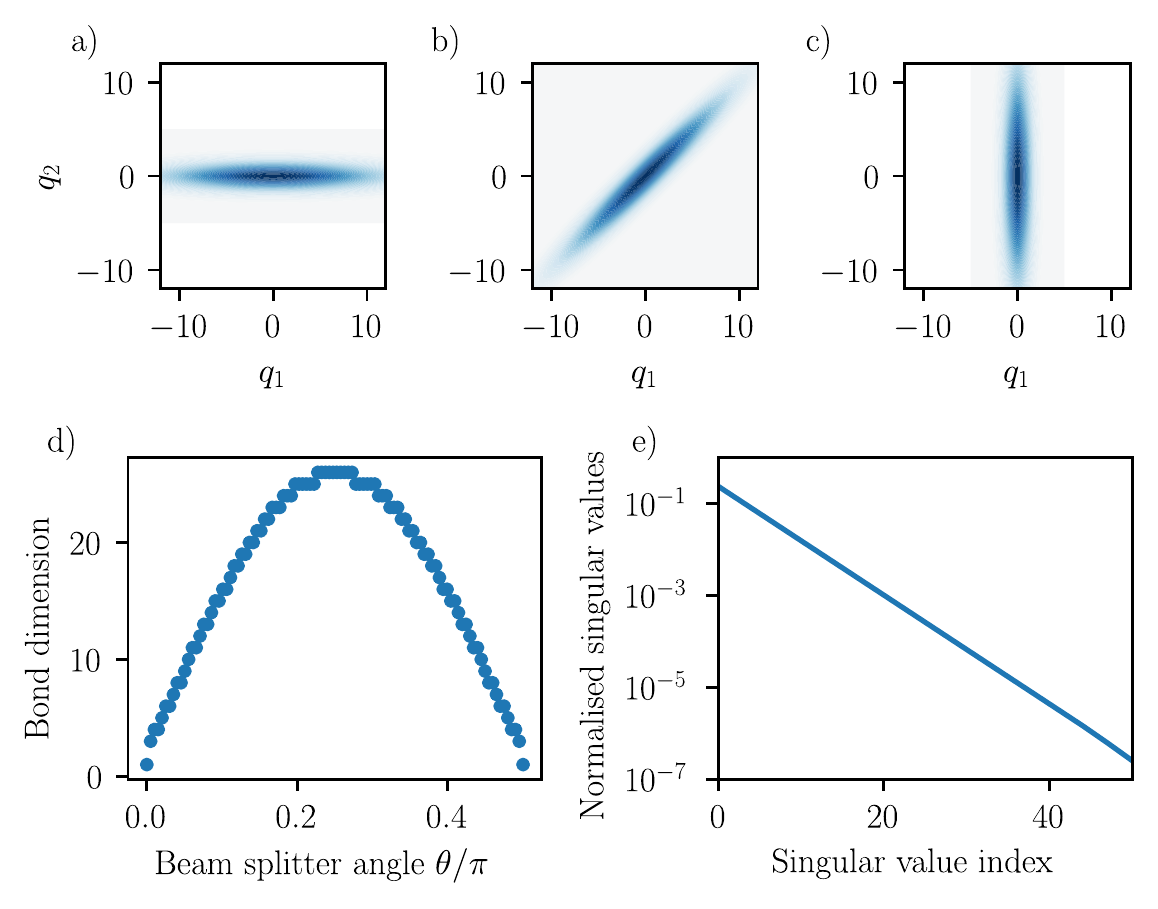}
    \caption{\textbf{(a)} To investigate the behavior of the FMPS representation, we consider a two-mode state with squeezing parameters $r_1 = 1, r_2 = -1$. We choose the two-dimensional wave function to be real-valued, and apply a beam splitting gate, effectively rotating the wave function in the $(q_1, q_2)$ space, as plotted in 
    for $\theta = 0, \pi/4$ and $\pi/2$. \textbf{(b)} Rotating the squeezed two-mode state on $N=200$ points in each dimension, we see that the bond dimension required to reach $0.999$ gate fidelity peaks at $\theta = \pi/4$. At either extreme the state reduces to a product state, corresponding to bond dimension 1. \textbf{(c)} The singular values at $\theta = \pi/4$ fall off exponentially fast, meaning that we can achieve a high fidelity approximation with low bond dimension.}
    \label{fig:total_squeeze}
\end{figure}

To build some intuition for the behavior of continuous variable states in the FMPS representation, we start by modeling a circuit of two squeezed modes with squeezing parameters $s_1=1, s_2=-1$ and a beam splitter with angle $\theta$:
\[
\Qcircuit @C=1em @R=.7em @!R {
\lstick{\ket{s_1}} & \multigate{1}{\text{BS}} & \qw\\
\lstick{\ket{s_2}} & \ghost{\text{BS}} & \qw\\
}
\]
For the purpose of this example we choose the representation to be a real valued position space wave function, with the beam splitter effectively rotating the state in the $(q_1,q_2)$ position space, see \cref{fig:total_squeeze}a-c. For strong squeezing, the $\theta= \pi/4$ rotation starts resembling a diagonal matrix (up to reordering), suggesting maximal entanglement, while the $\theta=0,\pi/2$ rotations approach rank one states. 
In \cref{fig:total_squeeze}d-e, we can indeed see that the bond dimension required for a $0.999$ fidelity of the truncated state peaks at $\theta = \pi/4$, corresponding to the maximal entanglement. Note that the bond dimension required to approximate the state is much smaller than the number of points used for discretisation, $N=200$.  Investigating the fidelity (in terms of singular value weight) of the maximally entangled state, we see that it falls off exponentially, indicating that the states can accurately be captured by a low bond dimension FMPS.

\begin{figure}
    \centering
    \includegraphics[width=\linewidth]{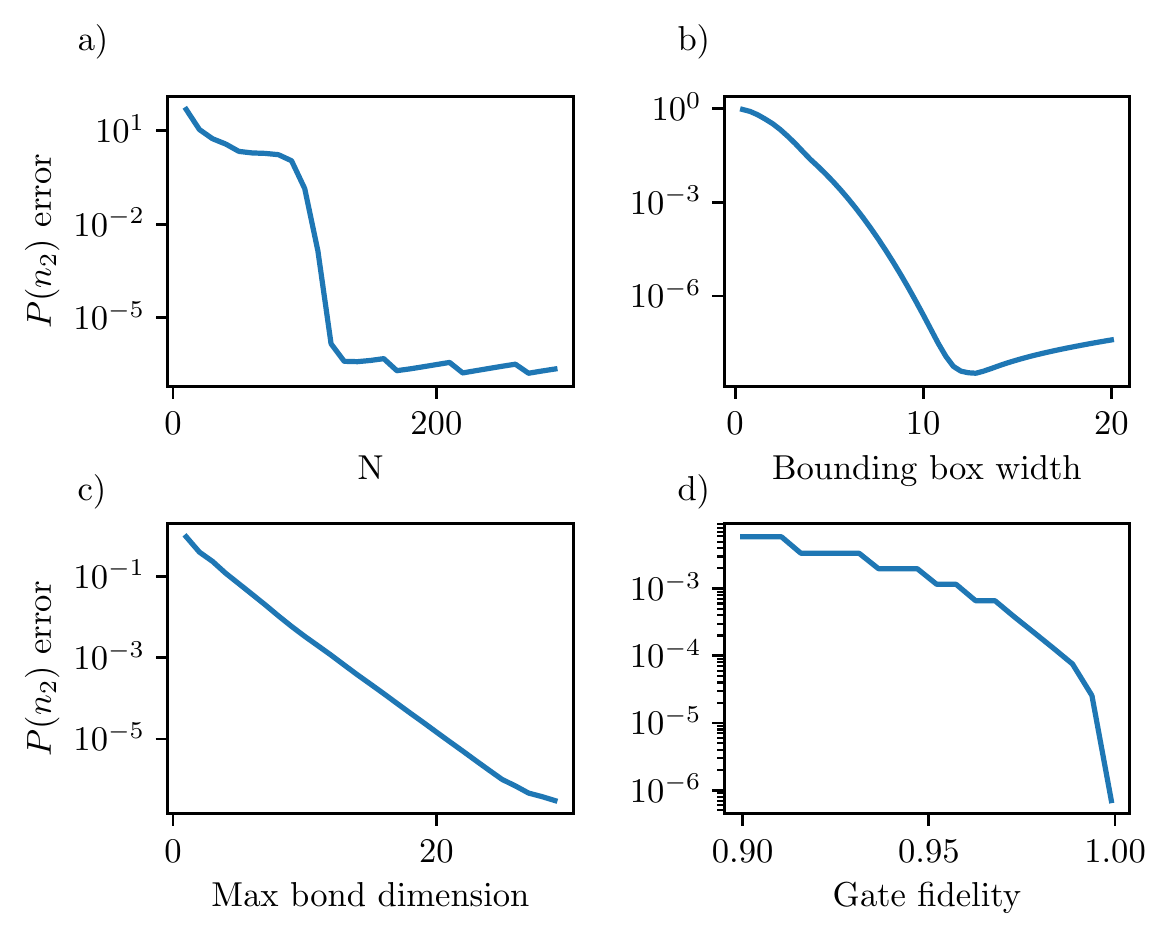}
    \caption{\textbf{(a)} The error of the FMPS method as compared to the analytical solution, see \cref{eq:Pn-error}, sees a sharp drop when increasing the number of grid points $N$, as also explored in \cref{sec:discretisation}. \textbf{(b)} Varying the width of the bounding box of the modes while keeping the resolution constant decreases the error as more of the state is captured, but will eventually increase the error if $N$ is not also increased. \textbf{(c)} Increasing the maximum  bond dimension of the FMPS reduces the error exponentially. \textbf{(d)} This can also be explored through the specification of a gate fidelity, which reduces the error in steps corresponding to increasing the bond dimension in integer steps.}
    \label{fig:n_error}
\end{figure}

The different sources of error (discretisation, bounded domain, rank reduction) can be explored systematically in this simple example. We consider the following setup where a squeezed state is beam split with the vacuum, and a photon number measurement is conducted in the second mode:
\[
\Qcircuit @C=1em @R=.7em @!R {
\lstick{\ket{s_1}} & \multigate{1}{\text{BS}} & \qw\\
\lstick{\ket{0}} & \ghost{\text{BS}} & \measureD{n_2}\\ 
}
\]
In this case, the probability of a given photon number measurement in the second mode can be expressed analytically as \cite{daknaGenerating1997}
\begin{multline}
    P_\text{true}(n_2) = \sqrt{\frac{1 - \kappa^2}{1 - \alpha^2}} 
\left( \frac{\alpha^2 (1 - |T|^2)}{|T|^2 (1 - \alpha^2)} \right)^{n_2}
\\
\times \sum_{k=0}^{\lfloor\frac{1}{2} n_2\rfloor} 
\frac{n_2!}{(n_2 - 2k)! (k!)^2 (2\alpha)^{2k}},
\end{multline}
where $T=\cos\theta$, $\kappa = \tanh(r_1)$ and $\alpha = T^2 |\kappa|$. We define the error
\begin{align}
    \epsilon_{n} = \sum_{n_2=0}^{n_{2,\text{max}}} \big|P_\text{true}(n_2) - P_\text{FMPS}(n_2)\big|,
    \label{eq:Pn-error}
\end{align}
and investigate it in \cref{fig:n_error} for (unless otherwise stated) $n_{2,max} = 50, \theta = \pi/4, N = 200$, a box width of $10$, gate fidelity of $0.999$, and max bond dimension set by gate fidelity. We see that we can directly control the sources of error, and that the error reduces (exponentially)  fast by tuning of each parameter independently.

\subsection{Cascading random beam splitters}
\label{sec:cascade}

As a second example, we consider a cascading beam splitter circuit with random rotation angles, and random input states $|\chi_j\rangle$. 
    \[
\Qcircuit @C=1em @R=.7em @!R {
\lstick{\ket{\chi_1}} & \multigate{1}{\text{BS}_1} &   \qw &   \qw \\
\lstick{\ket{\chi_2}} & \ghost{\text{BS}_1} & \multigate{1}{\text{BS}_2} & \qw \\
\lstick{\ket{\chi_3}} & \qw & \ghost{\text{BS}_2} & \qw & \\
& & & \ddots & \\
& & & & \multigate{1}{\text{BS}_{m-1}} & \qw \\
\lstick{\ket{\chi_m}} & \qw & \dots & & \ghost{\text{BS}_{m-1}} & \qw 
}
\] 
This represents a particularly clean implementation of the FMPS decomposition, as the 'entanglement depth' of the circuit is one. It can be experimentally realized in e.g. time-multiplexed systems, where a sequence of squeezed states are passed through a looped beam splitter and thus entangled with those states that passed through previously. This circuit implements repeated photon subtraction \cite{eatonNonGaussian2019,takaseGeneration2021,podoshvedovAlgorithm2023}, which is one suggested method of reliably producing non-Gaussian states, with cascaded circuits being of particular interest in generating GKP states from cat or squeezes states \cite{takaseGottesmanKitaevPreskill2023,konnoPropagating2024}. The cascaded quantum circuit is also known to exactly implement an MPS in the Fock basis, if the input states are number (Fock) states \cite{lubasch2018tensor}. Hence, we expect our FMPS simulations to perform particularly well in this scenario, independent of the input states. On the contrary, the Bosonic backend will perform poorly if the input states are non-Gaussian.

\begin{figure}
    \centering
    \includegraphics[width=\columnwidth]{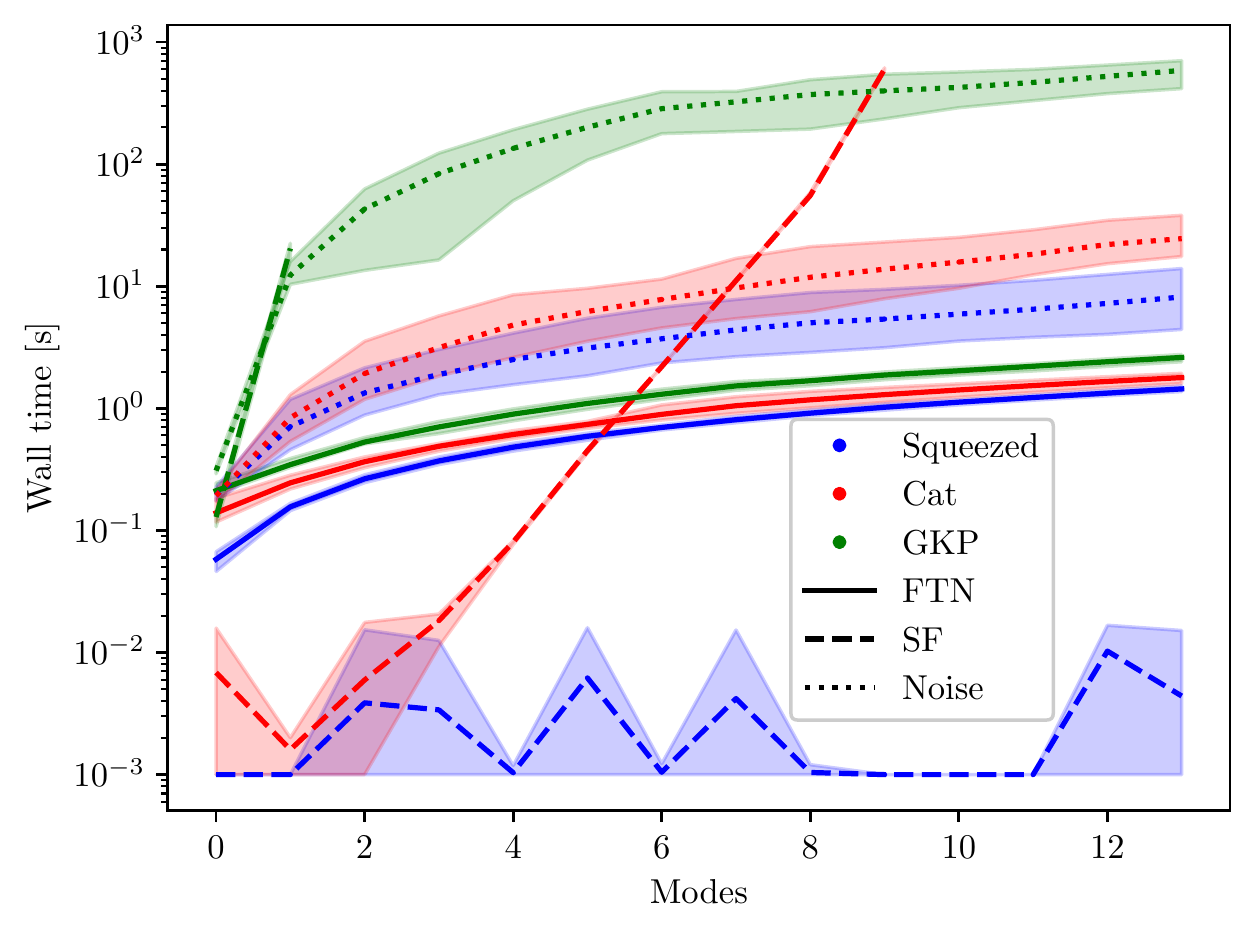}
    \caption{Simulating the cascading circuit, we see that the FMPS method (solid lines) scales sub-exponentially in the number of modes, independently of the input states. The SF method (dashed lines) in contrast scales exponentially for non-Gaussian states. This is to be expected, since the FMPS method scales with the total entanglement in the system, which saturates for shallow circuits. The dotted lines show the extra time taken to simulate the noise given by $\theta$ on the right hand side of \cref{fig:noise-diagram}. Adding noise significantly increases the simulation time of the cascading circuit, but the scaling remains sub-exponential.}
    \label{fig:cascade_comparison}
\end{figure}

\cref{fig:cascade_comparison} reports the simulation of the cascaded circuit with random rotation angles, and with random input states chosen from three families: (i) squeezed states with $|s|\leq 0.5$ (ii) Cat states with $|\alpha| \leq 2$, and (iii) GKP states with $\kappa = \Delta = 0.453$ (approximately corresponding to the default GKP state setting in SF). The phases ($\theta$ in Eqn. (\ref{eqn:cat}) and both phases $\theta$ and $\phi$ in Eqn. (\ref{eqn:gkp})) are sampled uniformly at random in $[0, 2\pi]$. 
The bond dimensions required for an accurate description saturates, causing the FMPS method's time complexity to grow sub-exponentially in the number of modes, for all three classes of input states. In contrast, the Bosonic backend for simulating CV circuits scales exponentially in the number of modes for non-gaussian input states..

The figure further shows the time added to the simulating using the method described in \cref{sec:noise} to simulate the random cascaded circuit with $10\%$ photon loss without parallelisation. This significantly increases the simulation time for the otherwise simple cascaded circuit, since the entanglement depth is increased. The scaling in the number of modes is however still polynomial. Adding noise in this way is immediately parallelisable, but the simulation has been left unparallelised to ensure a fair comparison with the SF library.

\subsection{Wide random circuits}

We now consider a more challenging benchmark for the FMPS method. A circuit of $l$ alternating layers of random beam splitters applied on $m$ modes in a 1D arrangement:
\[
\Qcircuit @C=1em @R=.7em @!R {
\lstick{\ket{\chi_0}} & \multigate{1}{\text{BS}} & \qw & \multigate{1}{\text{BS}} & \qw  \\
\lstick{\ket{\chi_1}} & \ghost{\text{BS}} & \multigate{1}{\text{BS}} & \ghost{\text{BS}} & \qw \\
\lstick{\ket{\chi_2}} & \multigate{1}{\text{BS}} & \ghost{\text{BS}} &\multigate{1}{\text{BS}}& \qw\\
\lstick{\ket{\chi_3}} & \ghost{\text{BS}} & \multigate{1}{\text{BS}} & \ghost{\text{BS}} & \qw\\
\lstick{\ket{\chi_4}} & \ghost{\text{BS}}  & \ghost{\text{BS}} & \ghost{\text{BS}} & \qw \\
& & \vdots &\\
&
}
\] 

\begin{figure}
    \centering
    \includegraphics[width=\columnwidth]{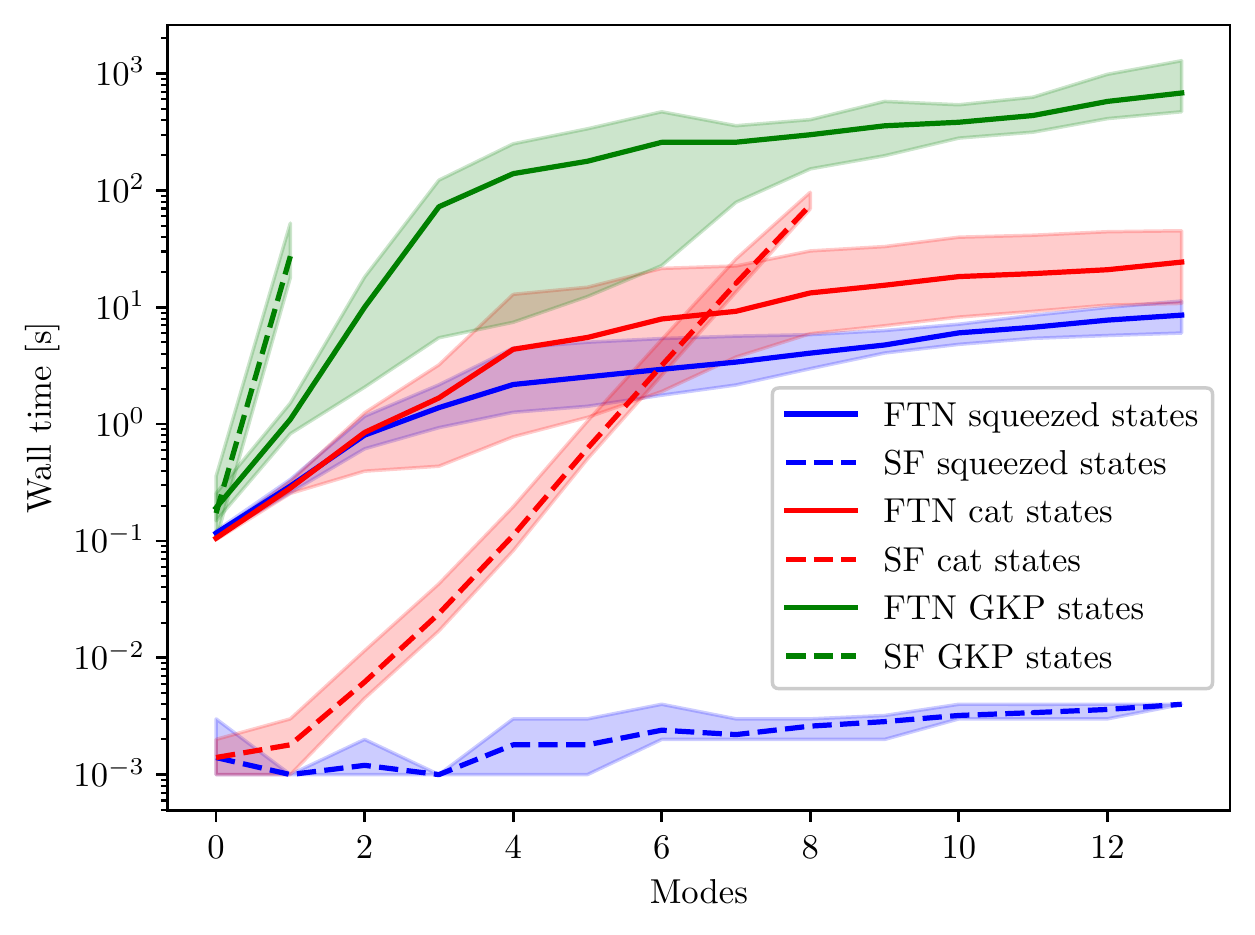}
    \caption{Fixing the wide circuit at 3 layers of beam splitters, we see that the bosonic backend of SF (dashed) always beats FMPS (solid) for squeezed states, that FMPS overtakes SF around 8 modes for cat states, and that FMPS is generally superior for GKP states. In all cases we see indications that FMPS scales sub-exponentially in the number of modes, as expected. Note that the SF cat and GKP simulations were stopped due to memory limitations.}
    \label{fig:3layer-wall-benchmark}
\end{figure}

Simulations of the circuits with $l=3$ and the settings given in \cref{tab:FMPS-wide-settings} are reported in \cref{fig:3layer-wall-benchmark}. Again, we see the same qualitative behavior as in the cascaded circuit, that the FMPS scales sub-exponentially in the number of modes. However, for highly non-Gaussian input states, the cost increases exponentially in the number of modes until setting into polynomial scaling. This is to account for the exponential scaling in the circuit depth. As with the cascaded circuit, the SF method scales exponentially in the number of modes for non-Gaussian input states.

\section{Conclusion}
We have introduced functional tensor networks for the simulation of continuous variable quantum circuits. This approach enables the efficient simulation of shallow circuits with many modes of highly non-Gaussian states, subject to general nearest-neighbor gate operations and measurements. The method leverages the strengths of Fock state descriptions for simulating non-Gaussian states and operations, along with the advantages of linear combinations of Gaussians for simulating states involving large photon numbers, such as GKP states. We have demonstrated how to implement the operations needed to implement the gates for universal CV quantum computing, as well as a series of useful measurements. In particular, we have shown that the method scales particularly favorable when modeling measurements. Finally, we have demonstrated how to simulate noise in the form of photon loss with an efficient strategy.

Through examples and benchmarks against state-of-the-art numerical methods we have shown that the functional tensor network method excels in simulating various systems that were previously infeasible to model at high precision. We expect the tool to be of significant value for the characterization of small scale quantum protocols involving bosonic qubits.  \\

\section{Acknowledgements}
The authors would like to thank Olga Solodovnikova, Jonas Neergaard-Nielsen and Nicola Pancotti for fruitful discussion. The work in this paper was supported by the Innovation Fund Denmark. FKM acknowledges support from Innovation Fund Denmark under grant no. 1063-00046B - “PhotoQ Photonic Quantum Computing”. ABM acknowledges support from the Novo Nordisk Foundation, Grant number NNF22SA0081175, NNF Quantum Computing Programme

\section{Data availability}
The data and code used for this paper can be found at \url{ https://doi.org/10.5281/zenodo.15162362}, along with guides for reproduction.

\bibliography{references}

@article{bigoniSpectral2016,
  title = {Spectral {{Tensor-Train Decomposition}}},
  author = {Bigoni, Daniele and {Engsig-Karup}, Allan P. and Marzouk, Youssef M.},
  year = {2016},
  month = jan,
  journal = {SIAM Journal on Scientific Computing},
  volume = {38},
  number = {4},
  pages = {A2405-A2439},
  publisher = {{Society for Industrial and Applied Mathematics}},
  issn = {1064-8275},
  doi = {10.1137/15M1036919}
}

@article{menicucci2006universal,
  title={Universal quantum computation with continuous-variable cluster states},
  author={Menicucci, Nicolas C and Van Loock, Peter and Gu, Mile and Weedbrook, Christian and Ralph, <? format?> Timothy C and Nielsen, Michael A},
  journal={Physical review letters},
  volume={97},
  number={11},
  pages={110501},
  year={2006},
  publisher={APS}
}

@article{putterman2025hardware,
  title={Hardware-efficient quantum error correction via concatenated bosonic qubits},
  author={Putterman, Harald and Noh, Kyungjoo and Hann, Connor T and MacCabe, Gregory S and Aghaeimeibodi, Shahriar and Patel, Rishi N and Lee, Menyoung and Jones, William M and Moradinejad, Hesam and Rodriguez, Roberto and others},
  journal={Nature},
  volume={638},
  number={8052},
  pages={927--934},
  year={2025},
  publisher={Nature Publishing Group UK London}
}

@article{lubasch2018tensor,
  title={Tensor network states in time-bin quantum optics},
  author={Lubasch, Michael and Valido, Antonio A and Renema, Jelmer J and Kolthammer, W Steven and Jaksch, Dieter and Kim, Myungshik S and Walmsley, Ian and Garc{\'\i}a-Patr{\'o}n, Ra{\'u}l},
  journal={Physical Review A},
  volume={97},
  number={6},
  pages={062304},
  year={2018},
  publisher={APS}
}

@article{liu2022validating,
  title = {Verifying Quantum Advantage Experiments with Multiple Amplitude Tensor Network Contraction},
  author = {Liu, Yong and Chen, Yaojian and Guo, Chu and Song, Jiawei and Shi, Xinmin and Gan, Lin and Wu, Wenzhao and Wu, Wei and Fu, Haohuan and Liu, Xin and Chen, Dexun and Zhao, Zhifeng and Yang, Guangwen and Gao, Jiangang},
  journal = {Phys. Rev. Lett.},
  volume = {132},
  issue = {3},
  pages = {030601},
  numpages = {6},
  year = {2024},
  month = {Jan},
  publisher = {American Physical Society},
  doi = {10.1103/PhysRevLett.132.030601},
  url = {https://link.aps.org/doi/10.1103/PhysRevLett.132.030601}
}

@article{pan2022simulation,
  title={Simulation of quantum circuits using the big-batch tensor network method},
  author={Pan, Feng and Zhang, Pan},
  journal={Physical Review Letters},
  volume={128},
  number={3},
  pages={030501},
  year={2022},
  publisher={APS}
}

@article{chabaud2023resources,
  title={Resources for bosonic quantum computational advantage},
  author={Chabaud, Ulysse and Walschaers, Mattia},
  journal={Physical Review Letters},
  volume={130},
  number={9},
  pages={090602},
  year={2023},
  publisher={APS}
}

@article{chabaud2021classical,
  title={Classical simulation of Gaussian quantum circuits with non-Gaussian input states},
  author={Chabaud, Ulysse and Ferrini, Giulia and Grosshans, Fr{\'e}d{\'e}ric and Markham, Damian},
  journal={Physical Review Research},
  volume={3},
  number={3},
  pages={033018},
  year={2021},
  publisher={APS}
}

@article{vinther2024variational,
  title = {Variational tensor network simulation of Gaussian boson sampling and beyond},
  author = {Vinther, Jonas and Kastoryano, Michael J.},
  journal = {Phys. Rev. A},
  volume = {112},
  issue = {2},
  pages = {022605},
  numpages = {12},
  year = {2025},
  month = {Aug},
  publisher = {American Physical Society},
  doi = {10.1103/z463-7gqy},
  url = {https://link.aps.org/doi/10.1103/z463-7gqy}
}

@article{oh2024classical,
  title={Classical algorithm for simulating experimental Gaussian boson sampling},
  author={Oh, Changhun and Liu, Minzhao and Alexeev, Yuri and Fefferman, Bill and Jiang, Liang},
  journal={Nature Physics},
  volume={20},
  number={9},
  pages={1461--1468},
  year={2024},
  publisher={Nature Publishing Group UK London}
}

@article{killoran2019strawberry,
  title={Strawberry fields: A software platform for photonic quantum computing},
  author={Killoran, Nathan and Izaac, Josh and Quesada, Nicol{\'a}s and Bergholm, Ville and Amy, Matthew and Weedbrook, Christian},
  journal={Quantum},
  volume={3},
  pages={129},
  year={2019},
  publisher={Verein zur F{\"o}rderung des Open Access Publizierens in den Quantenwissenschaften}
}

@article{bourassa2021fast,
  title={Fast simulation of bosonic qubits via Gaussian functions in phase space},
  author={Bourassa, J Eli and Quesada, Nicol{\'a}s and Tzitrin, Ilan and Sz{\'a}va, Antal and Isacsson, Theodor and Izaac, Josh and Sabapathy, Krishna Kumar and Dauphinais, Guillaume and Dhand, Ish},
  journal={PRX Quantum},
  volume={2},
  number={4},
  pages={040315},
  year={2021},
  publisher={APS}
}

@article{leghtas2013hardware,
  title={Hardware-efficient autonomous quantum memory protection},
  author={Leghtas, Zaki and Kirchmair, Gerhard and Vlastakis, Brian and Schoelkopf, Robert J and Devoret, Michel H and Mirrahimi, Mazyar},
  journal={Physical Review Letters},
  volume={111},
  number={12},
  pages={120501},
  year={2013},
  publisher={APS}
}

@article{girvin2023introduction,
  title={Introduction to quantum error correction and fault tolerance},
  author={Girvin, Steven M},
  journal={SciPost Physics Lecture Notes},
  pages={070},
  year={2023}
}

@article{ma2021quantum,
  title={Quantum control of bosonic modes with superconducting circuits},
  author={Ma, Wen-Long and Puri, Shruti and Schoelkopf, Robert J and Devoret, Michel H and Girvin, Steven M and Jiang, Liang},
  journal={Science Bulletin},
  volume={66},
  number={17},
  pages={1789--1805},
  year={2021},
  publisher={Elsevier}
}

@article{mirrahimi2014dynamically,
  title={Dynamically protected cat-qubits: a new paradigm for universal quantum computation},
  author={Mirrahimi, Mazyar and Leghtas, Zaki and Albert, Victor V and Touzard, Steven and Schoelkopf, Robert J and Jiang, Liang and Devoret, Michel H},
  journal={New Journal of Physics},
  volume={16},
  number={4},
  pages={045014},
  year={2014},
  publisher={IOP Publishing}
}

@article{gottesman2001encoding,
  title={Encoding a qubit in an oscillator},
  author={Gottesman, Daniel and Kitaev, Alexei and Preskill, John},
  journal={Physical Review A},
  volume={64},
  number={1},
  pages={012310},
  year={2001},
  publisher={APS}
}

@article{campagne2020quantum,
  title={Quantum error correction of a qubit encoded in grid states of an oscillator},
  author={Campagne-Ibarcq, Philippe and Eickbusch, Alec and Touzard, Steven and Zalys-Geller, Evan and Frattini, Nicholas E and Sivak, Volodymyr V and Reinhold, Philip and Puri, Shruti and Shankar, Shyam and Schoelkopf, Robert J and others},
  journal={Nature},
  volume={584},
  number={7821},
  pages={368--372},
  year={2020},
  publisher={Nature Publishing Group UK London}
}

@article{nathan2024self,
  title={Self-correcting GKP qubit and gates in a driven-dissipative circuit},
  author={Nathan, Frederik and O'Brien, Liam and Noh, Kyungjoo and Matheny, Matthew H and Grimsmo, Arne L and Jiang, Liang and Refael, Gil},
  journal={arXiv preprint arXiv:2405.05671},
  year={2024}
}

@article{chamberland2022building,
  title={Building a fault-tolerant quantum computer using concatenated cat codes},
  author={Chamberland, Christopher and Noh, Kyungjoo and Arrangoiz-Arriola, Patricio and Campbell, Earl T and Hann, Connor T and Iverson, Joseph and Putterman, Harald and Bohdanowicz, Thomas C and Flammia, Steven T and Keller, Andrew and others},
  journal={PRX Quantum},
  volume={3},
  number={1},
  pages={010329},
  year={2022},
  publisher={APS}
}

@article{bourassa2021blueprint,
  title={Blueprint for a scalable photonic fault-tolerant quantum computer},
  author={Bourassa, J Eli and Alexander, Rafael N and Vasmer, Michael and Patil, Ashlesha and Tzitrin, Ilan and Matsuura, Takaya and Su, Daiqin and Baragiola, Ben Q and Guha, Saikat and Dauphinais, Guillaume and others},
  journal={Quantum},
  volume={5},
  pages={392},
  year={2021},
  publisher={Verein zur F{\"o}rderung des Open Access Publizierens in den Quantenwissenschaften}
}

@article{takaseGeneration2021,
  title = {Generation of Optical {{Schr}}\"odinger Cat States by Generalized Photon Subtraction},
  author = {Takase, Kan and Yoshikawa, Jun-ichi and Asavanant, Warit and Endo, Mamoru and Furusawa, Akira},
  year = {2021},
  month = jan,
  journal = {Physical Review A},
  volume = {103},
  number = {1},
  pages = {013710},
  publisher = {{American Physical Society}},
  doi = {10.1103/PhysRevA.103.013710},
  urldate = {2023-04-13},
  
}

@article{larsen2021fault,
  title={Fault-tolerant continuous-variable measurement-based quantum computation architecture},
  author={Larsen, Mikkel V and Chamberland, Christopher and Noh, Kyungjoo and Neergaard-Nielsen, Jonas S and Andersen, Ulrik L},
  journal={Prx Quantum},
  volume={2},
  number={3},
  pages={030325},
  year={2021},
  publisher={APS}
}

@article{podoshvedovAlgorithm2023,
  title = {Algorithm of Quantum Engineering of Large-Amplitude High-Fidelity {{Schr\"odinger}} Cat States},
  author = {Podoshvedov, Mikhail S. and Podoshvedov, Sergey A. and Kulik, Sergei P.},
  year = {2023},
  month = mar,
  journal = {Scientific Reports},
  volume = {13},
  number = {1},
  pages = {3965},
  publisher = {{Nature Publishing Group}},
  issn = {2045-2322},
  doi = {10.1038/s41598-023-30218-6},
  urldate = {2023-05-11},
  copyright = {2023 The Author(s)},
  langid = {english},
}

@article{eatonNonGaussian2019,
  title = {Non-{{Gaussian}} and {{Gottesman}}\textendash{{Kitaev}}\textendash{{Preskill}} State Preparation by Photon Catalysis},
  author = {Eaton, Miller and Nehra, Rajveer and Pfister, Olivier},
  year = {2019},
  month = nov,
  journal = {New Journal of Physics},
  volume = {21},
  number = {11},
  pages = {113034},
  publisher = {{IOP Publishing}},
  issn = {1367-2630},
  doi = {10.1088/1367-2630/ab5330},
  urldate = {2023-08-24},
  langid = {english}
}

@article{weedbrookGaussian2012,
  title = {Gaussian Quantum Information},
  author = {Weedbrook, Christian and Pirandola, Stefano and {Garc{\'i}a-Patr{\'o}n}, Ra{\'u}l and Cerf, Nicolas J. and Ralph, Timothy C. and Shapiro, Jeffrey H. and Lloyd, Seth},
  year = {2012},
  month = may,
  journal = {Reviews of Modern Physics},
  volume = {84},
  number = {2},
  pages = {621--669},
  publisher = {{American Physical Society}},
  doi = {10.1103/RevModPhys.84.621},
  urldate = {2023-11-07}
}

@misc{olivaEntanglementaware2023,
  title = {An Entanglement-Aware Quantum Computer Simulation Algorithm},
  author = {Oliva, Maxime},
  year = {2023},
  month = jul,
  number = {arXiv:2307.16870},
  eprint = {2307.16870},
  primaryclass = {cond-mat, physics:quant-ph},
  publisher = {{arXiv}},
  doi = {10.48550/arXiv.2307.16870},
  urldate = {2023-10-26},
  archiveprefix = {arxiv}
}

@article{ayralDensityMatrix2023,
  title = {Density-{{Matrix Renormalization Group Algorithm}} for {{Simulating Quantum Circuits}} with a {{Finite Fidelity}}},
  author = {Ayral, Thomas and Louvet, Thibaud and Zhou, Yiqing and Lambert, Cyprien and Stoudenmire, E. Miles and Waintal, Xavier},
  year = {2023},
  month = apr,
  journal = {PRX Quantum},
  volume = {4},
  number = {2},
  pages = {020304},
  publisher = {{American Physical Society}},
  doi = {10.1103/PRXQuantum.4.020304},
  urldate = {2023-11-17}
}

@article{bourassaFast2021,
  title = {Fast {{Simulation}} of {{Bosonic Qubits}} via {{Gaussian Functions}} in {{Phase Space}}},
  author = {Bourassa, J. Eli and Quesada, Nicol{\'a}s and Tzitrin, Ilan and Sz{\'a}va, Antal and Isacsson, Theodor and Izaac, Josh and Sabapathy, Krishna Kumar and Dauphinais, Guillaume and Dhand, Ish},
  year = {2021},
  month = oct,
  journal = {PRX Quantum},
  volume = {2},
  number = {4},
  pages = {040315},
  publisher = {{American Physical Society}},
  doi = {10.1103/PRXQuantum.2.040315},
  urldate = {2022-11-29}
}

@article{vincentJet2022,
  title = {Jet: {{Fast}} Quantum Circuit Simulations with Parallel Task-Based Tensor-Network Contraction},
  shorttitle = {Jet},
  author = {Vincent, Trevor and O'Riordan, Lee J. and Andrenkov, Mikhail and Brown, Jack and Killoran, Nathan and Qi, Haoyu and Dhand, Ish},
  year = {2022},
  month = may,
  journal = {Quantum},
  volume = {6},
  pages = {709},
  publisher = {{Verein zur F\"orderung des Open Access Publizierens in den Quantenwissenschaften}},
  doi = {10.22331/q-2022-05-09-709},
  urldate = {2022-11-30},
  langid = {british}
}

@article{tzitrinProgress2020,
  title = {Progress towards Practical Qubit Computation Using Approximate {{Gottesman-Kitaev-Preskill}} Codes},
  author = {Tzitrin, Ilan and Bourassa, J. Eli and Menicucci, Nicolas C. and Sabapathy, Krishna Kumar},
  year = {2020},
  month = mar,
  journal = {Physical Review A},
  volume = {101},
  number = {3},
  pages = {032315},
  publisher = {{American Physical Society}},
  doi = {10.1103/PhysRevA.101.032315},
  urldate = {2023-09-04}
}

@book{sakurai_modern_2017,
	edition = {2},
	title = {Modern Quantum Mechanics:},
	isbn = {978-1-108-49999-6},
	url = {https://www.cambridge.org/core/product/identifier/9781108499996/type/book},
	shorttitle = {Modern Quantum Mechanics},
	publisher = {Cambridge University Press},
	author = {Sakurai, J. J. and Napolitano, Jim},
	urldate = {2022-09-16},
	date = {2017-09-21},
	doi = {10.1017/9781108499996},
    year = {2017},
}

@article{liuSimulating2023,
  title = {Simulating Lossy {{Gaussian}} Boson Sampling with Matrix-Product Operators},
  author = {Liu, Minzhao and Oh, Changhun and Liu, Junyu and Jiang, Liang and Alexeev, Yuri},
  year = {2023},
  month = nov,
  journal = {Physical Review A},
  volume = {108},
  number = {5},
  pages = {052604},
  publisher = {{American Physical Society}},
  doi = {10.1103/PhysRevA.108.052604},
  urldate = {2024-01-16}
}

@book{oppenheim_discrete-time_2010,
	location = {Upper Saddle River},
	edition = {3rd ed},
	title = {Discrete-time signal processing},
	isbn = {978-0-13-198842-2},
	pagetotal = {1108},
	publisher = {Pearson},
	author = {Oppenheim, Alan V. and Schafer, Ronald W.},
	date = {2010},
}

@article{shchesnovichNoise2019,
  title = {Noise in {{BosonSampling}} and the Threshold of Efficient Classical Simulatability},
  author = {Shchesnovich, Valery},
  year = {2019},
  month = jul,
  journal = {Physical Review A},
  volume = {100},
  number = {1},
  eprint = {1902.02258},
  eprinttype = {arxiv},
  primaryclass = {quant-ph},
  pages = {012340},
  issn = {2469-9926, 2469-9934},
  doi = {10.1103/PhysRevA.100.012340},
  archiveprefix = {arXiv}
}

@article{qiRegimes2020,
  title = {Regimes of {{Classical Simulability}} for {{Noisy Gaussian Boson Sampling}}},
  author = {Qi, Haoyu and Brod, Daniel J. and Quesada, Nicol{\'a}s and {Garc{\'i}a-Patr{\'o}n}, Ra{\'u}l},
  year = {2020},
  month = mar,
  journal = {Physical Review Letters},
  volume = {124},
  number = {10},
  pages = {100502},
  publisher = {{American Physical Society}},
  doi = {10.1103/PhysRevLett.124.100502}
}

@article{konnoPropagating2024,
  title = {Propagating {{Gottesman-Kitaev-Preskill}} States Encoded in an Optical Oscillator},
  author = {Konno, Shunya and Asavanant, Warit and Hanamura, Fumiya and Nagayoshi, Hironari and Fukui, Kosuke and Sakaguchi, Atsushi and Ide, Ryuhoh and China, Fumihiro and Yabuno, Masahiro and Miki, Shigehito and Terai, Hirotaka and Takase, Kan and Endo, Mamoru and Marek, Petr and Filip, Radim and {van Loock}, Peter and Furusawa, Akira},
  year = {2024},
  month = jan,
  journal = {Science},
  volume = {383},
  number = {6680},
  eprint = {2309.02306},
  primaryclass = {quant-ph},
  pages = {289--293},
  issn = {0036-8075, 1095-9203},
  doi = {10.1126/science.adk7560},
  urldate = {2024-06-24},
  archiveprefix = {arXiv}
}

@article{takaseGottesmanKitaevPreskill2023,
  title = {Gottesman-{{Kitaev-Preskill}} Qubit Synthesizer for Propagating Light},
  author = {Takase, Kan and Fukui, Kosuke and Kawasaki, Akito and Asavanant, Warit and Endo, Mamoru and Yoshikawa, Jun-ichi and {van Loock}, Peter and Furusawa, Akira},
  year = {2023},
  month = oct,
  journal = {npj Quantum Information},
  volume = {9},
  number = {1},
  pages = {1--11},
  publisher = {Nature Publishing Group},
  issn = {2056-6387},
  doi = {10.1038/s41534-023-00772-y},
  urldate = {2024-06-24},
  copyright = {2023 The Author(s)},
  langid = {english}
}

@article{daknaGenerating1997,
  title = {Generating Schr\"odinger-cat-like states by means of conditional measurements on a beam splitter},
  author = {Dakna, M. and Anhut, T. and Opatrn\'y, T. and Kn\"oll, L. and Welsch, D.-G.},
  journal = {Phys. Rev. A},
  volume = {55},
  issue = {4},
  pages = {3184--3194},
  numpages = {0},
  year = {1997},
  month = {Apr},
  publisher = {American Physical Society},
  doi = {10.1103/PhysRevA.55.3184},
  url = {https://link.aps.org/doi/10.1103/PhysRevA.55.3184}
}

@misc{marqversenImpact2025,
      title={Impact of finite squeezing on near-term quantum computations using GKP qubits}, 
      author={Frederik K. Marqversen and Andreas B. Michelsen and Janus H. Wesenberg and Nikolaj T. Zinner},
      year={2025},
      eprint={2507.15955},
      archivePrefix={arXiv},
      primaryClass={quant-ph},
      url={https://arxiv.org/abs/2507.15955}, 
}

\appendix

\section{First and second moment of phase rotated wave function}\label{app:moments}
We consider the phase rotation operator $\hat{R}(\theta) = e^{i\theta a^\dagger a}$ which enacts the following symplectic transformation:
\begin{equation}
    \hat{R}(\theta)^\dagger \begin{bmatrix}\hat{q} \\ \hat{p}\end{bmatrix} \hat{R}(\theta) = \begin{bmatrix} \cos\theta & -\sin\theta \\ \sin\theta & \cos\theta \end{bmatrix} \begin{bmatrix}\hat{q} \\ \hat{p}\end{bmatrix}
\end{equation}
Now consider the action of $\hat{R}$ on a quantum state
\begin{equation}
    \ket{\psi'} = R(\theta) \ket{\psi}
\end{equation}
The first moment of $\ket{\psi'}$ is then given by
\begin{gather}
    \langle q \rangle_{\psi'} = \bra{\psi'} \hat{q} \ket{\psi'}
    = \bra{\psi} \hat{R}(\theta)^\dagger \hat{q} \hat{R}(\theta) \ket{\psi}
        \\
    = \bra{\psi} (\cos\theta \hat{q} - \sin\theta \hat{p}) \ket{\psi}
\end{gather}
The result \cref{eq:first moment} now follows from inserting a resolution of identity $I = \int_\mathbb{R} \ket{q} \bra{q}$ and using the identities
\begin{equation}
    \bra{q} \hat{q} \ket{\psi} = q \psi(q)
        ,\quad
    \bra{q} \hat{p} \ket{\psi} = -i\pdv{}{q} \psi(q)
\end{equation}

Computing the second moment follows completely analogously from the observation
\begin{gather}
    \langle q^2 \rangle_{\psi'} = \bra{\psi'} \hat{q}^2 \ket{\psi'}
    = \bra{\psi} \left[ \hat{R}(\theta)^\dagger \hat{q} \hat{R}(\theta) \right]^2 \ket{\psi}
        \\
    = \bra{\psi} (\cos^2\theta \hat{q}^2 + \cos\theta \sin\theta (i - 2\hat{q}\hat{p}) + \sin^2\theta \hat{p}^2) \ket{\psi}
\end{gather}

\end{document}